\documentstyle{mn}
\input{psfig}
\newcommand{\etal}{{et al.~}}

\newcommand{\lta}{\la}
\newcommand{\gta}{\ga}

\newcommand{\kms}{\>{\rm km}\,{\rm s}^{-1}}

\newcommand{\Msun}{\>{\rm M_{\odot}}}

\newcommand{\dS}{\Delta S}
\newcommand{\dW}{\Delta \omega}
\newcommand{\dd}{\rm d}

\newcommand{\apj}{ApJ}
\newcommand{\apjs}{ApJS}
\newcommand{\aj}{AJ}
\newcommand{\mnras}{MNRAS}
\newcommand{\aap}{A\&A}

\newdimen\hssize
\newdimen\hdsize
\begin{document}


\title[The Universal Mass Accretion History of CDM Haloes]
      {The Universal Mass Accretion History of CDM Haloes}
\author[F.C. van den Bosch]
       {Frank C. van den Bosch\\
        Max-Planck Institut f\"ur Astrophysik, Karl Schwarzschild
         Str. 1, Postfach 1317, 85741 Garching, Germany}


\date{}

\pagerange{\pageref{firstpage}--\pageref{lastpage}}
\pubyear{2001}

\maketitle

\label{firstpage}


\begin{abstract}
  We  use the  extended Press-Schechter  formalism to  investigate the
  rate at which cold dark  matter haloes accrete mass.  We discuss the
  shortcomings of previous methods that  have been used to compute the
  mass  accretion histories  of  dark matter  haloes,  and present  an
  improved  method based on  the $N$-branch  merger tree  algorithm of
  Somerville \&  Kolatt.  We show  that this method no  longer suffers
  from  inconsistencies  in  halo  formation times,  and  compare  its
  predictions with high  resolution $N$-body simulations. Although the
  overall  agreement is reasonable,  there are  slight inconsistencies
  which  are most easily  interpreted as  a reflection  of ellipsoidal
  collapse  (as   opposed  to   spherical  collapse  assumed   in  the
  Press-Schechter formalism). We show  that the average mass accretion
  histories  follow a  simple,  universal profile,  and  we present  a
  simple  recipe  for  computing  the two  scale-parameters  which  is
  applicable to a wide range of halo masses and cosmologies.  Together
  with  the universal profiles  for the  density and  angular momentum
  distributions  of   CDM  haloes,  these   universal  mass  accretion
  histories provide  a simple but accurate framework  for modeling the
  structure and formation of  dark matter haloes.  In particular, they
  can be  used as  a backbone for  modeling various aspects  of galaxy
  formation where  one is  not interested in  the detailed  effects of
  merging. As an  example we use the universal  mass accretion history
  to compute the rate at  which dark matter haloes accrete mass, which
  we compare to the cosmic star formation history of the Universe.
\end{abstract}


\begin{keywords}
cosmology: theory ---
galaxies: formation ---
galaxies: halos ---
stars: formation ---
dark matter.
\end{keywords}


\section{Introduction}
\label{sec:intro}

In the standard  Cold Dark Matter (CDM) family  of cosmological models
dark  matter  haloes form  hierarchically  through  the accretion  and
merging of smaller structures that  condense out of a Gaussian initial
density field.   The rate  at which these  dark matter haloes  grow in
mass sets, amongst others, the rate  at which baryons can cool to form
luminous objects.  Therefore, knowledge of the mass  accretion rate of
dark  matter haloes  is an  essential ingredient  in  our cosmological
framework of structure formation.

Since  the  collapse  and virialization   of  dark matter  haloes is a
non-linear process,   one  generally resorts  to   numerical  $N$-body
simulations to  study the formation and  evolution of structures  in a
CDM Universe.    However, this method    has  a number  of   important
drawbacks.  First of   all, numerical simulations are  computationally
expensive,  making  it  unfeasible  to   explore    a wide range    of
cosmological models.  Secondly,  computational limitations only  allow
simulations of  small volumes and/or   a  small dynamical mass  range.
This  makes  it virtually  impossible    to simultaneously follow  the
formation and evolution of objects  from sub-galactic scale up to  the
scale of (super)clusters.

An  important  alternative is  provided  by  the Press-Schechter  (PS)
formalism  (Press  \&  Schechter  1974)  which allows  a  much  faster
exploration  of halo  mass accretion  histories  for a  wide range  of
cosmologies  and masses.   In addition,  the PS  formalism  provides a
framework that allows us to  gain insights into the physical processes
involved. The main {\it ansatz} of the PS-formalism is to consider the
initial density field extrapolated  {\it linearly} to redshift $z$ and
smoothed  on some  typical mass  scale $M$.   Regions above  a certain
threshold value are then associated with collapsed objects of mass $M$
at  redshift $z$.   Motivated by  Birkhoff's theorem  it  is therefore
assumed  that  the  non-linear  evolution  of  density  perturbations,
described by means of a spherical `Top-Hat' model (Gunn \& Gott 1972),
does not influence the remainder  of the Universe.  Press \& Schechter
(1974) used this formalism to compute the mass function of dark matter
haloes  as  function  of redshift,  which  has  been  found to  be  in
remarkably  good  agreement  with  results from  $N$-body  simulations
(e.g., Efstathiou \etal 1988; Carlberg \& Couchman 1989; Lacey \& Cole
1994; Gelb  \& Bertschinger 1994; Ma 1996).   

The  PS  theory  has  been  extended to  give  the  {\it  conditional}
probabilities  $P(M_2,z_2 \vert M_1,  z_1)$ that  a given  particle at
redshift $z_1$  inside a halo of  mass $M_1$ at an  earlier time $z_2$
was embedded  in a halo  of mass $M_2$  (Bond \etal 1991;  Bower 1991;
Lacey \&  Cole 1993).   This extended Press-Schechter  (hereafter EPS)
formalism  is   easily  manipulated   to  yield  merger   rates,  halo
formation/survival times, and various  other statistics (Lacey \& Cole
1993;  hereafter  LC93).   Of   particular  importance  has  been  the
construction  of  halo  merger  trees  (e.g.,  Cole  \&  Kaiser  1988;
Kauffmann \&  White 1993; Somerville  \& Kolatt 1999; Sheth  \& Lemson
1999),  which  are  widely  used  in semi-analytical  models  for  the
formation  of galaxies  (e.g.,  Kauffmann, White  \& Guiderdoni  1993;
Somerville \& Primack 1999; Cole \etal 1994, 2000).

In some cases, however, one  is not interested in knowing the detailed
distribution of halo  progenitor masses, but one merely  wants to know
the rate at which the halo  mass increases with time. For instance, in
the disk galaxy formation models  of Firmani \& Avila-Reese (2000) and
van den Bosch (2001a,b) the  assumption is explicitly made that haloes
accrete  their mass  smoothly; i.e.,  one ignores  the fact  that mass
accretion involves the  merging of progenitor haloes.  In  the case of
disk galaxies this simplification  is permitted since the fragility of
disks suggest that mergers have not played an important role.  In this
paper we  therefore use  the EPS formalism  to construct  average mass
accretion  histories of  dark matter  haloes, which  we define  as the
ensemble average $\langle M(z)/M_0  \rangle$.  Here $M(z)$ is the halo
mass as function of redshift and  $M_0$ is the present day mass of the
halo.  The main motivation for  this study is to investigate whether a
simple,  universal   form  exists  for   $\langle  M(z)/M_0  \rangle$.
Together  with the  universal  profiles for  the density  distribution
(Navarro,  Frenk \& White  1997) and  angular momentum  (Bullock \etal
2001) of CDM haloes, such  universal mass accretion history provides a
complete  description of the  structure and  evolution of  dark matter
haloes, which can be used as  a framework for detailed modeling of the
formation  of galaxies. In  addition, it  is to  be expected  that the
accretion  history of  dark matter  haloes is  directly linked  to the
cosmic  star formation  history.  A  universal mass  accretion history
might  therefore proof  useful  in trying  to  understand the  rapidly
improving  observations of  the star  formation rates  as  function of
redshift.

This paper  is organized as  follows.  In Section~\ref{sec:background}
we  start  with  a  brief  description of  the  PS-formalism  and  its
extension   based  on   the   excursion  set   formalism.   Next,   in
Section~\ref{sec:models} we describe  an improved method for computing
mass  accretion  histories,  which  we test  against  high  resolution
$N$-body      simulations     in      Section~\ref{sec:nbody}.      In
Section~\ref{sec:universal}  we  derive  a simple,  universal  fitting
formula  for  the average  mass  accretion  histories  of dark  matter
haloes,  which  is applicable  to  a wide  range  of  halo masses  and
cosmologies.   In Section~\ref{sec:appl}  we discuss  a  comparison of
star  formation and  halo mass  accretion  rates, and  we conclude  in
Section~\ref{sec:concl}.  A  step-by-step  recipe  for  computing  the
average mass  accretion history for a  dark matter halo  of given mass
and in a given cosmology is presented in Appendix~A.

\section{Theoretical background}
\label{sec:background}

In  the  standard model  for  structure  formation  the density  field
$\delta({\bf x}) = \rho({\bf x})/\bar{\rho} - 1$ is considered to be a
Gaussian random field, which  is therefore completely specified by the
power spectrum  $P(k)$.  As long as  $\delta \ll 1$ the  growth of the
perturbations is linear and $\delta({\bf x},t_2) = \delta({\bf x},t_1)
D(t_2)/D(t_1)$,  where  $D(t)$  is  the linear  growth  factor.   Once
$\delta({\bf x})$ exceeds a critical threshold $\delta^{0}_{\rm crit}$
non-linear effects become important and the perturbation will start to
collapse to form a virialized object (halo). In what follows we define
$\delta_0$ as  the initial density field linearly  extrapolated to the
present time.  In terms of  $\delta_0$, regions that have collapsed to
form virialized objects at redshift $z$ are then associated with those
regions  for  which  $\delta_0  > \delta_c(z)  \equiv  \delta^{0}_{\rm
crit}/D(z)$.\footnote{Here  $D(z)$ corresponds  to  the linear  growth
factor normalized to unity at the present.} 

In order to assign masses to these collapsed regions, the PS formalism
considers the density field  $\delta_0$ smoothed with a spatial window
function (filter) $W(r;R_f)$, where  $R_f$ is a characteristic size of
the filter.  There is a considerable  amount of freedom  in choosing a
window  function, and  here we  adopt the  often used  spatial top-hat
filter
\begin{equation}
\label{TH}
W(r;R_f) = \left\{\begin{array}{ll}
3/(4 \pi R_f^3) & \mbox{$(r \leq R_f)$} \\
0               & \mbox{$(r > R_f)$}
\end{array} 
\right.
\end{equation}
The main advantage  of this filter over for  example a Gaussian filter
or a $k$-space top-hat filter is that it is straightforward to compute
the mass contained  within the window function: $M  = 4 \pi \bar{\rho}
R_f^3/3$,  with $\bar{\rho}$ the  mean mass  density of  the Universe.
The {\it ansatz} of the PS  formalism is that the probability that the
density field  smoothed with  $W(r;R_f)$ exceeds $\delta_c(z)$  is the
same as  the fraction  of mass  that at redshift  $z$ is  contained in
haloes with  masses greater than $M$.  This results in  the well known
unconstrained  PS  mass function  for  the  comoving  mass density  of
haloes:
\begin{eqnarray}
\label{PS}
\lefteqn{{{\dd}n \over  {\dd} \, {\rm  ln} \, M}(M,z) \, {\dd}M =} \nonumber \\ 
& & \sqrt{2 \over  \pi} \, \bar{\rho} \, {\delta_c(z) \over
\sigma^2(M)} \, \left| {{\dd} \sigma \over  {\dd}  M}\right|  
\,  {\rm  exp}\left[-{\delta_c^2(z)  \over  2 \sigma^2(M)}\right] \, {\dd}M
\end{eqnarray}
(Press \& Schechter 1974). Here  $\sigma^2(M)$ is the mass variance of
the smoothed density field given by
\begin{equation}
\label{variance}
\sigma^2(M) = {1 \over 2 \pi^2} \int_{0}^{\infty} P(k) \;
\widehat{W}^2(k;R_f) \; k^2 \; {\dd}k.
\end{equation}
with $\widehat{W}(k;R_f)$  the Fourier transform  of $W(r;R_f)$, which
for the spatial top-hat filter used here is given by:
\begin{equation}
\label{THfour}
\widehat{W}(k;R_f) = {3 \over (k R_f)^3} \left[ \sin(k R_f) - k R_f 
\cos(k R_f)\right]
\end{equation}

The extended Press-Schechter model  developed by Bond \etal (1991), is
based on the  excursion set formalism.  For each  point one constructs
`trajectories'  $\delta(M)$  of  the  linear  density  field  at  that
position as  function of the smoothing  mass $M$.  In  what follows we
adopt the notation of LC93 and use the variables $S = \sigma^2(M)$ and
$\omega =  \delta_c(z)$ to label  mass and redshift,  respectively. In
the limit  $R_f \rightarrow  \infty$ one has  that $(S,\omega)=(0,0)$,
which  can  be considered  the  starting  point  of the  trajectories.
Increasing  $S$ corresponds  to decreasing  the filter  mass  $M$, and
$\delta(S)$  starts  to wander  away  from  zero,  executing a  random
walk. The fraction of matter in collapsed objects in the mass interval
$M$, $M+{\rm d}M$ at redshift  $z$ is now associated with the fraction
of  trajectories that have  their {\it  first upcrossing}  through the
barrier  $\omega =  \delta_c(z)$ in  the interval  $S$,  $S+{\rm d}S$,
which is given by
\begin{equation}
\label{probS}
P(S ,\omega) \; {\dd}S = {1  \over \sqrt{2 \pi}} \; 
{\omega  \over S^{3/2}} \; 
{\rm exp}\left[-{\omega^2 \over 2 S}\right] \; {\dd}S
\end{equation}
(Bond  \etal 1991;  Bower  1991; LC93).   After  conversion to  number
counting,  this probability function  yields the  PS mass  function of
equation~(\ref{PS})

Since for random walks the upcrossing probabilities are independent of
the path  taken    (i.e., the upcrossing  is  a   Markov process), the
probability for a change $\dS$ in a time step $\dW$ is simply given by
equation~(\ref{probS}) with $S$  and $\omega$ replaced with $\dS$  and
$\dW$, respectively. This allows one to  immediate write down the {\it
conditional}  probability that a  particle in a halo  of mass $M_2$ at
$z_2$ was embedded in a halo of mass $M_1$ at $z_1$ (with $z_1 > z_2$)
as
\begin{eqnarray}
\label{probSS}
\lefteqn{P(S_1,\omega_1 \vert  S_2,\omega_2) \; {\dd}S_1 =} \nonumber \\
& & {1  \over \sqrt{2 \pi}} \;
{(\omega_1    -   \omega_2)    \over   (S_1    -    S_2)^{3/2}} \; {\rm
exp}\left[-{(\omega_1 - \omega_2)^2 \over 2 (S_1 - S_2)}\right] \;
{\dd}S_1 
\end{eqnarray}
Converting from  mass weighting to  number weighting, one  obtains the
average number  of progenitors  at $z_1$ in  the mass  interval $M_1$,
$M_1 + {\rm d}M_1$ which by  redshift $z_2$ have merged to form a halo
of mass $M_2$:
\begin{eqnarray}
\label{condprobM}
\lefteqn{{{\dd}N \over {\dd}M_1}(M_1,z_1 \vert M_2,z_2) \; {\dd}M_1 =} 
\nonumber \\
& & {M_2 \over
M_1} \; P(S_1,\omega_1 \vert S_2,\omega_2) \; 
\left\vert {{\dd}S \over {\dd M}} \right\vert \; {\dd}M_1.
\end{eqnarray}

\section{Constructing Mass Accretion Histories}
\label{sec:models}
 
Using the EPS conditional probabilities for halo progenitor masses one
can construct detailed  histories of the mass assembly  of dark matter
haloes.   Here we  are interested  in computing  the  ``mass accretion
histories'' (hereafter MAHs) of dark matter haloes, which we define as
$\Psi(M_0,z) \equiv M(z)/M_0$.  Here $M(z)$ is defined as  the mass of
the ``main progenitor halo'', and $M_0$ is the halo mass at $z=0$.

When  tracing  back  in time  each  halo  breaks  up  in a  number  of
progenitor  haloes, which  themselves  break up  in progenitors,  etc.
Given this complicated history of  the mass evolution of a halo (often
referred  to  as the  ``merger-tree''),  we  need  to define  what  we
actually mean  with ``the  main progenitor'' at  an earlier  time.  We
follow  previous studies  (LC93; Eisenstein  \& Loeb  1996;  Nusser \&
Sheth 1999)  and define $M(z)$ as  the main trunk of  the merger tree,
i.e., at each time step we  associate $M(z)$ with the mass of the most
massive progenitor (hereafter MMP), and we follow that progenitor, and
that  progenitor only,  further back  in  time.  This  way the  ``main
progenitor halo''  never actually accretes other haloes  that are more
massive than  itself.  Note that  although at each branching  point we
follow the most  massive branch, this does not  necessarily imply that
the main progenitor is also  the most massive of {\it all} progenitors
at a given redshift.
\begin{figure*}
\centerline{\psfig{figure=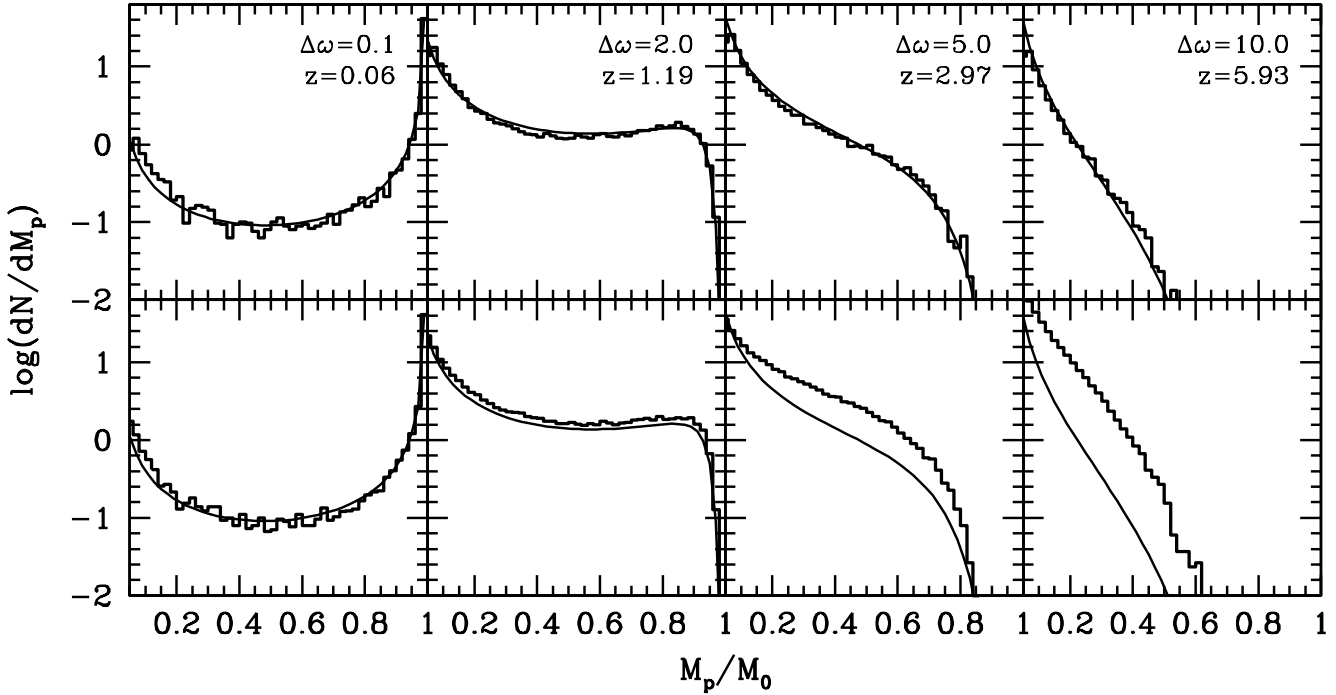,width=\hdsize}}
\caption{Mass  functions of progenitors  (by number)  for a  halo with
present day mass $M_0 = 10^{12}  h^{-1} \Msun$ in an EdS universe with
$h=0.65$  and $\sigma_8=1.0$.   Histograms correspond  to  the average
number of progenitors found in 15000 random realizations of the merger
tree. Only  progenitors more massive  than five percent of  the parent
are  considered (i.e.,  $M_{\rm  min} =  0.05  M$, see  text), and  we
therefore  only plot  results for  $M_p/M_0 >  0.05$. The  solid lines
correspond to the  EPS prediction given by equation~(\ref{condprobM}).
Results are  shown for  four `output' times  of the merger  trees, the
corresponding $\Delta \omega$ and redshift of which are indicated.  In
the  upper  panels  results  are  shown  for  merger  trees  that  are
constructed using  a fixed  time step of  $\Delta \omega =  0.1$. This
results in  progenitor numbers in excellent agreement  with the direct
EPS prediction.  In the lower  panels, progenitor masses  are computed
using a  single time  step (with $\Delta  \omega$ as indicated  in the
upper panels). In  this case, too many progenitors  are found compared
to  the EPS prediction.  This indicates  that for  the merger  tree to
yield self-consistent results sufficiently small time steps have to be
used.}
\label{fig:massfunc}
\end{figure*}

In  what follows we  use $M_0$  to denote  the present  day mass  of a
halo. Each individual time step we use $M$ to refer to the parent mass
for  which we seek  the most  massive progenitor  a time  step $\Delta
\omega$ earlier, and we use $M_p$ to indicate the mass of a progenitor
halo.

\subsection{Previous methods}
\label{sec:previous}

In  an infinitesimally small  time interval  $\Delta \omega$  a change
$\Delta  S$  results  from  a  {\it  single}  merger.   Therefore,  an
approximate  method for constructing  MAHs is  to consider  small time
steps,  and to assume  that the  change in  mass associated  with that
finite time interval reflects a single merger event. In that case, the
MMP has  a mass $M_p  \geq M/2$ and  the construction of  MAHs becomes
very simple:  each time step  one draws a  single $\Delta S$  from the
probability distribution
\begin{equation}
\label{probdS}
P(\dS ,\dW)  \; {\dd}\dS = {1 \over \sqrt{2  \pi}} \; {\dW \over
\dS^{3/2}} \; {\rm exp}\left[-{(\dW^2) \over 2 \dS}\right] \; {\dd}\dS
\end{equation}
and  one   defines  the   mass  of  the   main  progenitor   as  ${\rm
max}(M_p,M-M_p)$,  where  $\sigma^2(M_p)  =  \sigma^2(M) +  \dS$.   An
alternative approach which leads to  almost similar results is to only
accept     values      of     $\Delta     S$      in     the     range
$[0,\sigma^2(M/2)-\sigma^2(M)]$.   This  method,  which  we  call  the
`binary'  method,  was  suggested  by  LC93,  and  has  been  used  by
Eisenstein \& Loeb (1996) to  compute the minimum intrinsic scatter in
the Tully-Fisher relation.  However,  as already pointed by LC93, this
method leads  to some inconsistencies regarding  halo formation times,
the  reason for  which is  easily understood.   The number  density of
progenitor masses diverges  at small mass (a direct  reflection of the
fact that for CDM $\sigma^2(M) \rightarrow \infty$ when $M \rightarrow
0$),  and  the  assumption  of  a  single  merger  event  brakes  down
dramatically for finite  time steps, even when chosen  very small.  In
other words,  there is a finite, non-negligible,  probability that the
mass of  the MMP is  less than $M/2$.   This introduces, at  each time
step, a systematic bias towards a main progenitor that is too massive,
resulting in formation redshifts that are too high.

An alternative method for constructing MAHs was suggested by Nusser \&
Sheth  (1999), who  draw the  progenitor  mass from  the {\it  number}
weighted distribution  function (equation~[\ref{condprobM}]) with $M/2
\leq  M_p \leq  M$.   Although, as  they  show, this  leads to  better
consistency  with halo  formation times,  this method  suffers  from a
similar shortcoming  as again the assumption  is made that  the MMP is
always more massive than $M/2$. However, this is only formally true in
the limit  $\Delta \omega  \rightarrow 0$, for  which the  integral of
${\dd}N/{\dd}M_1$  (equation~[\ref{condprobM}])   from  $M/2$  to  $M$
becomes unity (i.e.,  it is certain that the MMP  is more massive than
$M/2$).

\subsection{An improved method}
\label{sec:new}

The discussion  above suggests that  in order to construct  MAHs using
finite time  steps, one  has to drop  the assumption of  single merger
events.  This  implies that  each time step  one needs to  construct a
complete  set of  halo  masses $M_i$  that  are to  be considered  the
progenitor haloes that a time step $\dW$ later have merged to form the
parent halo with mass $M$. An important constraint on the set $M_i$ is
that $\sum_i  M_i = M$, i.e, the  total mass needs to  be conserved at
each time step. The MAH is then easily constructed by picking the most
massive of  $M_i$, and repeating  the same procedure stepping  back in
time until the mass of the main progenitor is as small as desired.

Unfortunately,  the  construction of  sets  of  progenitors  is not  a
trivial matter.  In  addition to conserving mass at  each time step, a
successful merger  tree also has  to satisfy the requirement  that, at
each  time step,  the distribution  of  the number  of progenitors  as
function        of        mass        is        consistent        with
equation~(\ref{condprobM}).  However,  since   the  number  of  haloes
diverges  at  very  small  masses,  one must  impose  a  minimum  halo
mass. Progenitors with masses  below this threshold are not considered
as individual  haloes, but their  mass is assumed to  accrete smoothly
onto the parent halo.  As  was previously pointed out by Somerville \&
Kolatt  (1999; hereafter  SK99), there  seems to  be no  algorithm for
drawing  sets of  progenitor  masses that  satisfies both  constraints
simultaneously.  Kauffmann  \& White (1993)  circumvented this problem
by reproducing  the progenitor mass distribution exactly,  but by only
enforcing mass  conservation approximately. Here we  follow the scheme
of  SK99,  which  conserves  mass  exactly  while  only  approximately
reproducing the progenitor mass distribution.  This method, termed the
$N$-branch method with  accretion, has been shown to  yield results in
good  agreement with  numerical simulations  (Somerville  \etal 2000).
Here  we briefly  summarize the  method  and we  refer the  interested
reader to SK99 for more details  (as well as for a detailed discussion
on other, less successful methods for constructing merger trees).

The  method  of  SK99  is  based  on  drawing  halo  masses  from  the
mass-weighted probability function~(\ref{probdS}).  With each new halo
drawn it is  checked whether the sum of  the progenitor masses exceeds
the mass of the parent $M$.  If  this is the case the halo is rejected
and a new progenitor mass is drawn.  Any progenitor with $M_p < M_{\rm
min}$ is added to the  mass component $M_{\rm acc}$ that is considered
to  be  accreted  onto the  parent  in  a  smooth fashion  (i.e.,  the
formation  history of  these small  mass progenitors  is  not followed
further back in time). Here $M_{\rm min}$ is a free parameter that has
to be chosen sufficiently small.  This procedure is repeated until the
total mass  left ($M - M_{\rm acc}  - \sum M_p$) is  less than $M_{\rm
min}$. This remaining mass is  assigned to $M_{\rm acc}$ and one moves
on to the next time step.
\begin{figure*}
\centerline{\psfig{figure=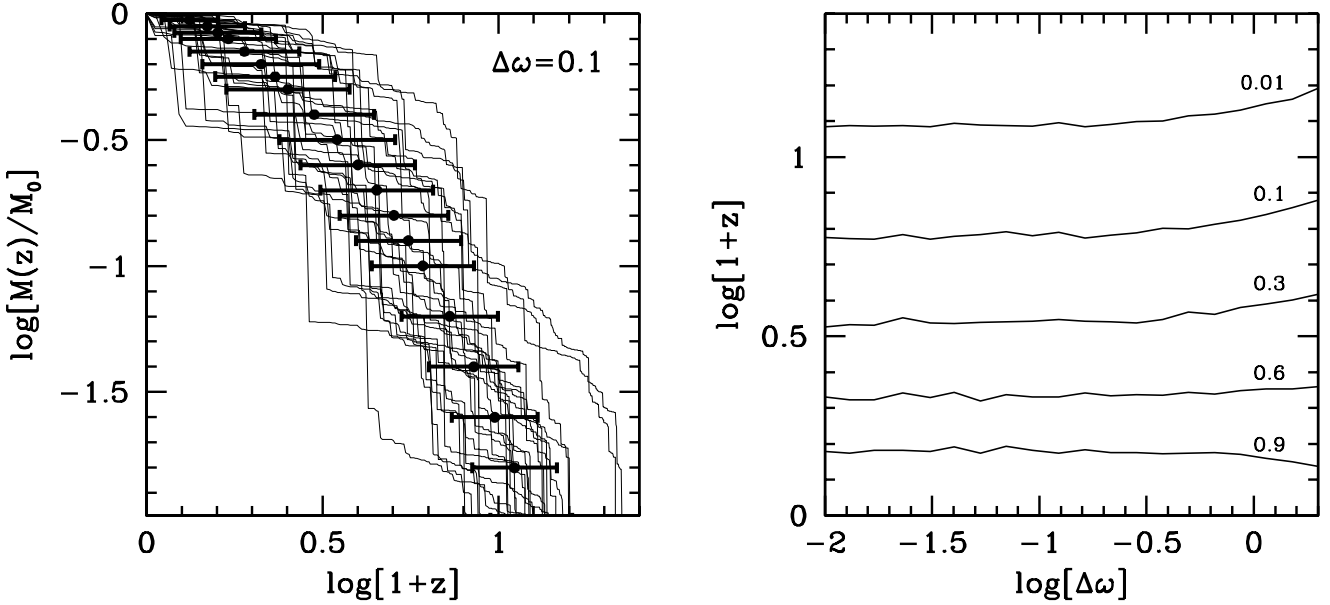,width=\hdsize}}
\caption{The panel  on the left plots  25 random MAHs for  a halo with
$M_0  = 5.0  \times  10^{11} h^{-1}  \Msun$  in an  EdS universe  with
$h=0.65$  and  $\sigma_8=1.0$. The  thick  solid  dots with  errorbars
correspond to  the average and standard deviations  as determined from
1000 such realizations. A time step  of $\Delta \omega = 0.1$ is used.
The panel on the right plots curves of constant $\langle \Psi \rangle$
(as indicated)  as function of the  time step $\Delta  \omega$ used in
the  construction  of  the  MAHs.   Each  $\langle  \Psi  \rangle$  is
determined  from  1000  random  realizations  of  the  mass  accretion
history.  For  $\Delta \omega  < 0.3$ the  average MAHs converge  to a
robust  result, while  for larger  time steps  $\langle  \Psi \rangle$
depends  on the  actual  value  of $\Delta  \omega$  used. This  again
suggests   that  a   successful  construction   of  MAHs   requires  a
sufficiently  small time  step  and  we adopt  $\Delta  \omega =  0.1$
throughout.}
\label{fig:example}
\end{figure*}

In principle, since the  upcrossing of trajectories through a boundary
is a  Markov process,  the statistics of  progenitor masses  should be
independent of the  time steps taken.  However, the  SK99 algorithm is
based on the  {\it single} halo probability (equation~[\ref{probdS}]),
which  does  not  contain  any  information about  the  {\it  set}  of
progenitors  that  make up  the  mass  of  $M$ (mass  conservation  is
enforced `by  hand', by rejecting progenitor masses  that overflow the
mass budget).  Therefore it is  not clear whether the results are time
step independent, and  this has to be tested.  In  the upper panels of
Figure~\ref{fig:massfunc}   we  plot   the   number  distribution   of
progenitor masses at various  redshifts obtained using the SK99 scheme
with a fixed  time step of $\Delta \omega =  0.1$. Results are plotted
for a halo with $M_0 =  10^{12} h^{-1} \Msun$ in an Einstein-de Sitter
(EdS)  universe   with  $\Omega_0   =  1$,  $\Omega_{\Lambda}   =  0$,
$\sigma_8=1.0$,   and   $h=0.65$.    Histograms  correspond   to   the
distributions obtained from $15000$  random realizations of the merger
tree, with $M_{\rm min} = 0.05  M$.  Solid lines correspond to the EPS
prediction of equation~(\ref{condprobM}).  In the lower panels we show
the  same results,  except that  we have  now computed  the progenitor
masses in a {\it single} time step (as indicated in the upper panels);
i.e., in  the panels on the  right, the upper panel  shows the results
when  using  100  time steps  of  $\Delta  \omega  = 0.1$,  each  time
computing the progenitors of  all previous progenitors etc.  The lower
panel, on the other hand, computed the progenitors using a single time
step  of $\Delta  \omega =  10.0$. As  can be  seen, when  using small
enough  time steps  the  number  density of  progenitor  masses is  in
excellent agreement with equation~(\ref{condprobM}). However, when too
large  time steps  are used,  the method  over-predicts the  number of
progenitors  quite  dramatically.  We  thus  conclude, confirming  the
results of SK99, that as long as small enough time steps are used, the
algorithm outlined above provides  an accurate method for constructing
merger trees.

Based on this scheme we use the following algorithm to construct MAHs:  
\begin{description}
\item[(1)] Choose a present day halo  mass $M_0$ and set $M = M_0$ and
$z = 0$.

\item[(2)] Set $M_{\rm left} = M$ and  compute the progenitor redshift
$z_p$ from $\dW = \delta_c(z_p) - \delta_c(z)$.

\item[(3)] Draw $\dS$ from the probability distribution~(\ref{probdS})
and   compute   the   corresponding   progenitor   mass   $M_p$   from
$\sigma^2(M_p) = \sigma^2(M) + \dS$.
 
\item[(4)] If  $M_p >  M_{\rm left}$ the  progenitor mass is  too big:
goto 3

\item[(5)]   Set  the   mass   of  the   MMP   to  $M_{\rm   MMP}={\rm
max}[M_p,M_{\rm MMP}]$ and the mass left in the set to $M_{\rm left} =
M_{\rm left} - M_p$.

\item[(6)] If $M_{\rm  MMP} \geq M_{\rm left}$  then we have found the
MMP.  In that case we proceed to the next  time step: we set $M(z_p) =
M_{\rm MMP}$, $M = M_{\rm MMP}$, $z=z_p$, and goto 2.

\item[(7)] Goto 3

\end{description}
This procedure is repeated until the mass of the main progenitor is as
small as desired. Note that, as is evident from step~5, we do not need
to construct an  entire set of progenitors; we can  stop once the most
massive of the progenitors already  drawn is larger than the mass left
in the  set, and we  thus do not  need to define a  minimum progenitor
mass  $M_{\rm  min}$.   Throughout  we  use  a  fitting  function  for
$\sigma^2(M)$ which is accurate to  better than $0.5$ percent over the
entire  mass range  $10^{6} h^{-1}  \Msun \leq  M \leq  10^{16} h^{-1}
\Msun$ (see Appendix~A). The power spectrum $P(k)$ is characterized by
the shape parameter $\Gamma$. Unless  stated otherwise we use the form
suggested by Sugiyama (1995)
\begin{equation}
\label{gamma}
\Gamma = \Omega_0 h \; {\rm exp}\left[ -\Omega_b (1 + \sqrt{2
h}/\Omega_0) \right]
\end{equation}
and  we adopt  a  baryon mass  density  of $\Omega_b  = 0.019  h^{-2}$
(Tytler \etal 1999).

We  now define  the {\it  average} mass  accretion  history (hereafter
AMAH) of a halo of mass $M_0$ as
\begin{equation}
\label{aMAH}
\langle \Psi(M_0,z) \rangle = 
{1 \over N} \sum_{i=1}^{N} \Psi_i(M_0,z)
\end{equation}
where the  summation is  over an ensemble  of $N$  random realizations
$\Psi_i(M_0,z)$.   Unless   stated  otherwise  we   use  $N=1000$  and
$\dW=0.1$.  In  the left panel of Figure~\ref{fig:example}  we show an
example.  The thin  lines are 25 random realizations for  the MAH of a
halo with  $M_0 = 5  \times 10^{11} h^{-1}  \Msun$ in an  EdS universe
with  $\sigma_8=1.0$  and $h=0.65$.   The  solid  dots with  errorbars
correspond  to the  AMAH  and its  standard  deviation (averaged  over
$1000$ random realizations).

The  right   panel  of  Figure~\ref{fig:example}   plots  the  average
redshifts at which $\langle \Psi \rangle = (0.01, 0.1, 0.3, 0.6, 0.9)$
as function  of the time step  $\dW$.  Note how  only for sufficiently
small time steps ($\dW \lta 0.3$) the values of $\langle \Psi \rangle$
are  not time  step  dependent.  This  reflects  the problem  outlined
above,  that  if  the  time  step  is too  big,  the  distribution  of
progenitor masses is  no longer consistent with the  EPS prediction of
equation~(\ref{condprobM}).   Results   for  other  halo   masses  and
cosmologies  are   similar,  and  we  therefore  adopt   $\dW  =  0.1$
throughout.

\section{Comparison with numerical simulations}
\label{sec:nbody}

Although the (extended)  PS formalism is a valuable  tool in the study
of  hierarchical structure  formation,  it  is based  on  a number  of
questionable assumptions (e.g., spherical collapse).  Therefore, it is
essential  that any  statistic  extracted using  the  PS formalism  is
tested  against  numerical simulations. 
\begin{figure}
\centerline{\psfig{figure=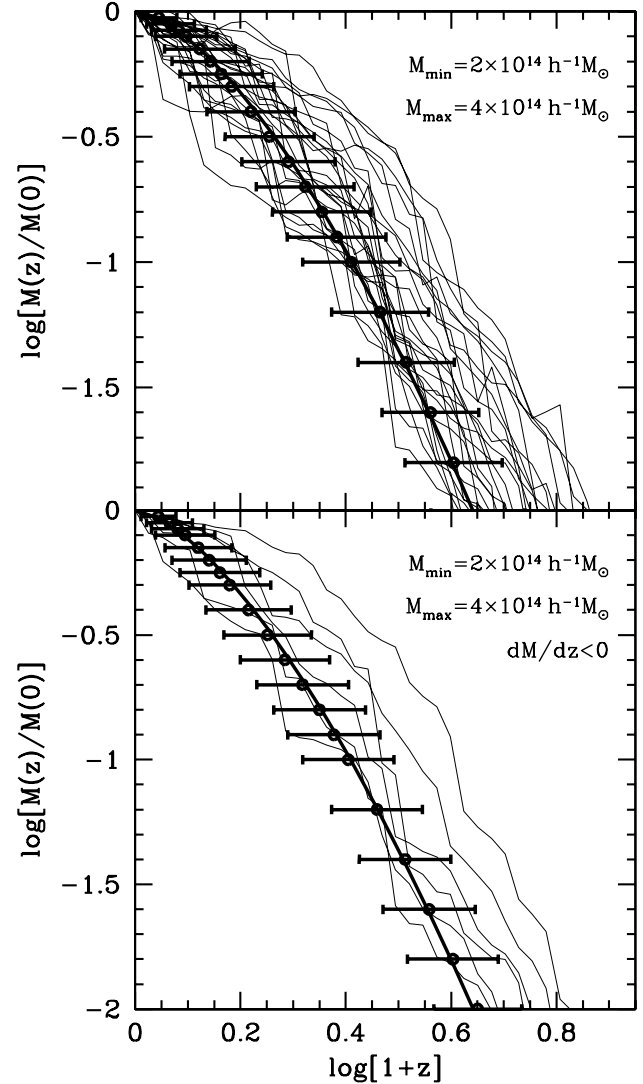,width=\hssize}}
\caption{The thin lines represent  mass accretion histories for haloes
in the $\Lambda$CDM GIF simulation.  In upper panel plots the MAHs for
$25$ (randomly selected) haloes  with $2.0 \times 10^{14} h^{-1} \Msun
\leq M_{\rm  FOF} \leq  4.0 \times 10^{14}  h^{-1} \Msun$.   The thick
lines and dots  with errorbars correspond to the  average MAH computed
using our method based on the EPS  formalism for a halo mass of $M_0 =
2.75 \times  10^{14} h^{-1} \Msun$,  corresponding to the  average FOF
mass of the 25 haloes in the simulation. As can be seen, haloes in the
simulation seem to  form a little bit earlier as  predicted by the EPS
formalism.  In the lower panel we plot the MAHs of those FOF groups in
the above mass  interval which have ${\dd}M/{\dd}z <  0$ at all times;
only 6 haloes obey these criteria.   As can be seen, the MAHs of these
haloes  are not  statistically  different from  those  in the  central
panel,  and   equally  inconsistent  with  the   EPS  prediction  (see
discussion in text).}
\label{fig:gifmah}
\end{figure}

\subsection{Description of the simulations}
\label{sec:simdes}

In order  to test our algorithm  for constructing MAHs  we compare our
results with  the high resolution  $\Lambda$CDM simulation of  the GIF
project (Kauffmann \etal 1999;  Diaferio \etal 2001).  This simulation
is   performed   using   the   parallel   adaptive   particle-particle
particle-mesh (AP$^3$M) code called  Hydra (Couchman, Thomas \& Pearce
1995) and follows  the evolution of $256^3$ ($\sim  16.8 \times 10^6$)
particles  in a  $141.3 h^{-1}$Mpc  box.  The  cosmological parameters
used  are  $\Omega_0  =  0.3$,  $\Omega_{\Lambda}=0.7$,  $h=0.7$,  and
$\Gamma=0.21$.    The  power-spectrum   of  density   fluctuations  is
normalized to $\sigma_8=0.9$, in agreement with the observed abundance
of  massive clusters  (Eke, Cole  \& Frenk  1996).  Particle  mass and
softening  length  are  $1.4  \times  10^{10} h^{-1}  \Msun$  and  $30
h^{-1}$kpc, respectively.

Particle positions and velocities are stored at 43 different redshifts
between $z=12$ and $z=0$.  At  each output time, haloes are identified
using  the  standard  friends-of-friends (FOF)  percolation  algorithm
(Davis  \etal 1985),  where a  linking length  of 0.2  times  the mean
inter-particle separation is adopted. The mass associated with each of
these haloes, denoted by $M_{\rm  FOF}$, is simply given by the number
of particles linked together times the single particle mass.

A  halo at redshift  $z_2$ is  defined as  a progenitor  of a  halo at
redshift $z_1 < z_2$ if (i)  more than half its particles are included
in  the halo  at  $z_1$, and  (ii)  its most  bound  particle is  also
included. Using lists of progenitors for all haloes and for all output
times\footnote{Available at http://www.mpa-garching.mpg.de/GIF/} a MAH
is constructed as  follows.  We start with a halo  at z=0 and identify
all its  progenitors at  the previous time  step $z_1$.  The  new halo
mass $M(z_1)$ is  then simply defined as the mass  of the most massive
of  these  progenitors,  and   this  procedure  is  repeated  for  all
subsequent  output   times.  A   more  detailed  description   of  the
construction of the  halo merger trees from these  GIF simulations can
be found in Kauffmann \etal (1999).

\subsection{Comparison of mass accretion histories}
\label{sec:simmah}

Here we  compare the MAHs of  dark matter haloes in  the simulation to
the   AMAHs  computed  using   the  EPS   formalism  as   outlined  in
Section~\ref{sec:new}.

In the upper panel of  Figure~\ref{fig:gifmah} we plot the MAHs for 25
(randomly  selected) haloes  in  the simulation  which  at $z=0$  have
masses in the range $2.0  \times 10^{14} h^{-1} \Msun \leq M_{\rm FOF}
\leq 4.0 \times  10^{14} h^{-1} \Msun$.  This mass  range is chosen to
ensure haloes  with large  numbers of particles,  so that we  are less
sensitive to resolution  issues.  We only accept haloes  for which the
MAHs can be traced back to  the point where $\Psi \leq 0.01$ (i.e., in
some cases no  progenitors can be identified in  the simulation before
the mass  of the main progenitor  has fallen below one  percent of the
present day mass).  The thick  solid dots with errorbars correspond to
the  AMAH computed  using  the  EPS formalism  (for  exactly the  same
cosmological parameters  as used  in the simulation)  for a  halo with
$M_0  = 2.75  \times 10^{14}  h^{-1} \Msun$  which corresponds  to the
average  mass of  the $25$  haloes  in the  simulation.  Although  the
overall  shape of  the  MAHs and  the  amounts of  scatter are  fairly
similar,  the MAHs  of  the  haloes in  the  numerical simulation  are
systematically  offset to higher  redshifts (i.e.,  the haloes  in the
simulation seem, on average, to form somewhat earlier).

One possible explanation for this  inconsistency is that the haloes in
the simulation often  have progenitors that are more  massive than the
parent.  This can  have several causes. Haloes in  the simulations are
susceptible to tidal  stripping, which can cause the  mass of the main
progenitor to actually {\it decrease} with time. In addition, during a
(high  velocity)  encounter  two  haloes  may  temporarily  be  linked
together by  the FOF  algorithm, which causes  $M(z)$ to  increase and
subsequently decrease again. These effects  are not modeled by the EPS
formalism, which  only allows halo masses  to grow with  time.  We can
investigate the effect this has on  the statistics of our MAHs by only
selecting haloes  from the  simulation that, at  each time  step, have
${\dd}M/{\dd}z < 0$. Only 6 of the 35 haloes in our mass interval obey
this  criterion, and  their MAHs  are plotted  in the  lower  panel of
Figure~\ref{fig:gifmah}. Although  the number statistics  are poor, it
is evident that these MAHs  are not statistically different from those
that occasionally  show ${\dd}M/{\dd}z >  0$, and they show  a similar
inconsistency  with respect to  the expected  AMAH.  We  thus conclude
that  the inconsistencies between  the MAHs  from the  simulations and
those from  the EPS  formalism are  not related to  the fact  that the
latter does not  take possible mass loss into  account.  A more likely
cause      for     the      inconsistency     is      discussed     in
Section~\ref{sec:ellipsoid}.

\subsection{Comparison of halo formation times}
\label{sec:simform}

Another useful  statistic for comparison  is the distribution  of halo
formation redshifts.   We use  the definition of  LC93 and  define the
halo formation  redshift, $z_f$, as  the redshift where $\Psi  = 0.5$,
i.e., where  the mass of the  main progenitor is half  the present day
mass.

Lacey   \& Cole   (1993)  presented two    methods for computing   the
distribution  of formation times.  The   first is based  on the number
weighted     mass     distribution    of       progenitor       haloes
(equation~[\ref{condprobM}]).  As   argued by LC93,  integrating  this
equation   from   $M_0/2$ to $M_0$  gives    the  probability that the
progenitor mass is  more massive than $M_0/2$, which  is equal  to the
probability that  the halo formation time  was earlier than this. Upon
defining the scaled variables
\begin{equation}
\label{tildeS}
\tilde{S} = {\sigma^2(M) - \sigma^2(M_0) \over 
\sigma^2(M_0/2) - \sigma^2(M_0)}
\end{equation}
and
\begin{equation}
\label{omegaf}
\tilde{\omega}_f = {\delta_c(z_f) - \delta_c(0) \over 
\sqrt{\sigma^2(M_0/2) - \sigma^2(M_0)}}
\end{equation}
one can  write the probability  distribution for halo  formation times
as:
\begin{equation}
\label{probomegaf}
P(\tilde{\omega}_f) = {1 \over \sqrt{2 \pi}} \int_0^1 {M_0 \over
M}  \left({\tilde{\omega}_f^2 \over  \tilde{S}^{5/2}}  - {1  \over
\tilde{S}^{3/2}} \right) {\rm exp}\left[-{\tilde{\omega}_f^2 \over
2 \tilde{S}}\right] {\dd}\tilde{S},
\end{equation}
where  $M$ is  solved from  equation~(\ref{tildeS}). The  advantage of
using  the  variables   $\tilde{S}$  and  $\tilde{\omega}_f$  is  that
$P(\tilde{\omega}_f)$ depends  only very mildly on  mass and cosmology
(this dependence is largely absorbed by the variables themselves). 

Another method for computing halo formation times is to use the actual
MAHs  themselves, and  to  identify  the redshift  at  which the  main
progenitor mass  equals half  the present day  mass.  LC93  used their
`binary'  Monte-Carlo  method  for  constructing  MAHs  (discussed  in
Section~\ref{sec:previous}),     and      found     a     distribution
$P(\tilde{\omega}_f)$    that     was    offset    from     that    of
equation~(\ref{probomegaf})  to higher  values  of $\tilde{\omega}_f$.
Lacey \&  Cole (1994) determined  the distributions of  halo formation
times   of  dark   matter  haloes   in  simulations   with  scale-free
power-spectra, and  found them to  be in excellent agreement  with the
analytical prediction of equation~(\ref{probomegaf}), but inconsistent
with $P(\tilde{\omega}_f)$  derived using the  Monte-Carlo method.  As
indicated  in  Section~\ref{sec:previous},  the  `binary'  Monte-Carlo
method  used by  LC93 is  based  on the  false premise  that the  most
massive progenitor has  mass $M_p > M/2$, which  leads to a systematic
offset of halo formation redshifts to too high values.
\begin{figure*}
\centerline{\psfig{figure=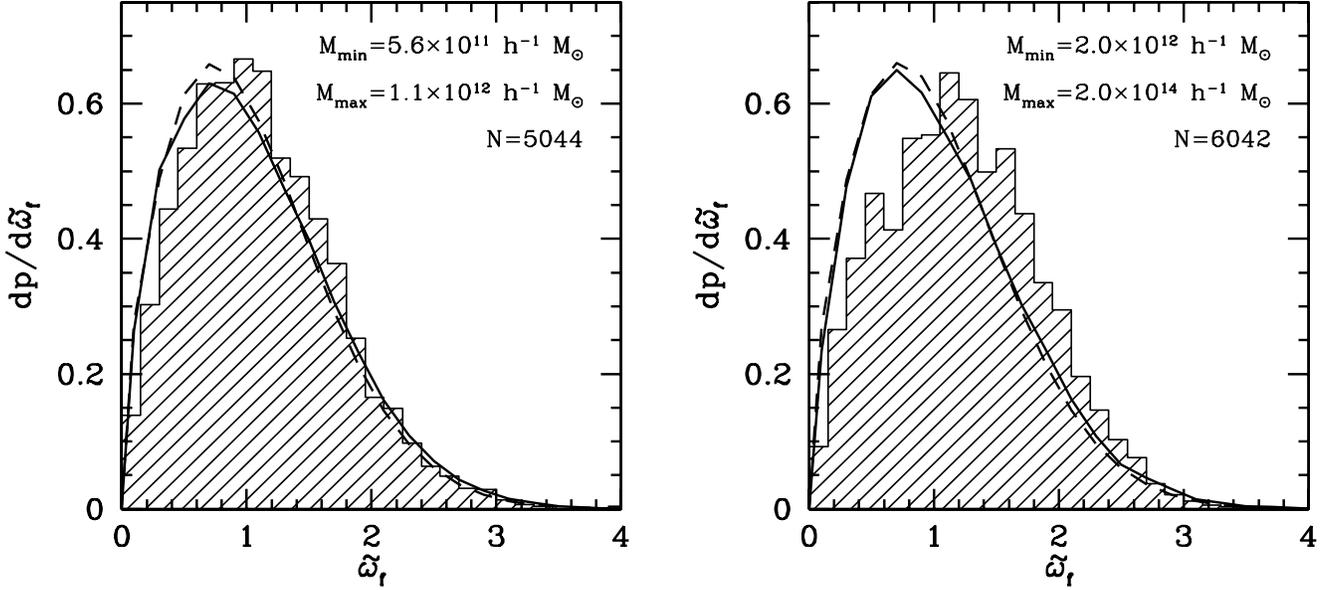,width=\hdsize}}
\caption{The hatched histograms correspond to the distribution of halo
formation    times    (parameterized    by   the    scaled    variable
$\tilde{\omega}_f$) for  haloes in the simulation (with  FOF masses as
indicated in the two panels). The solid lines are the distributions of
$\tilde{\omega}_f$ computed  from $2 \times  10^5$ random realizations
of the  MAH for a  halo with $M_0  = 8.4 \times 10^{11}  h^{-1} \Msun$
(left  panel) and  $M_0  =  1.1 \times  10^{13}  h^{-1} \Msun$  (right
panel). These masses correspond to  the average halo masses in the two
simulation samples.  Dashed  lines correspond to $P(\tilde{\omega}_f)$
of  equation~(\ref{probomegaf})  for  the  same  masses,  and  are  in
excellent agreement with the  distributions obtained directly from the
MAHs.  However, the formation times  in the simulation are offset from
those of the EPS formalism, whereby the discrepancy is larger for more
massive haloes.}
\label{fig:histogif}
\end{figure*}

We  now  re-examine  the  issue  of halo  formation  times  using  the
high-resolution  $\Lambda$CDM simulation and  our improved  method for
constructing  MAHs.   In  order  to  improve accuracy  we  use  linear
interpolation between the time steps that bracket $\Psi=0.5$ (for both
the  haloes in the  simulations as  well as  for the  MAHs constructed
using  the EPS  formalism). We  convert $z_f$  to the  scaled variable
$\tilde{\omega}_f$,  so that  $P(\tilde{\omega}_f)$ depends  only very
weakly  on  cosmology  and  halo  mass.  This  allows  us  to  compare
distributions   of  $\tilde{\omega}_f$   for  relatively   large  mass
intervals, providing better statistics.

We compute  the distribution of $\tilde{\omega}_f$ for  two samples of
haloes.  The  first consists  of 5044 haloes  with present  day masses
$5.6 \times  10^{11} h^{-1}  \Msun \leq M_{\rm  FOF} \leq  1.12 \times
10^{12} h^{-1}  \Msun$).  This mass range corresponds  to haloes that,
at  $z=0$, consist  of  between  40 and  80  particles. The  resulting
distribution  of  $\tilde{\omega}_f$   is  indicated  by  the  hatched
histogram in  the left panel of  Figure~\ref{fig:histogif}.  The solid
curve  corresponds to the  distribution of  $\tilde{\omega}_f$ derived
from $2  \times 10^5$ MAHs for a  halo with $M_0 =  8.4 \times 10^{11}
h^{-1} \Msun$ (corresponding to the average mass of the 5044 haloes in
the  simulation), using  the same  cosmological parameters  as  in the
simulation.  The two distributions are in good agreement, although the
mean  for   the  simulated  haloes   is  slightly  offset   to  higher
$\tilde{\omega}_f$.      The    dashed     curve     corresponds    to
$P(\tilde{\omega}_f)$ of equation~(\ref{probomegaf}),  also for a mass
of  $8.4  \times  10^{11}  h^{-1}  \Msun$.  This  distribution  is  in
excellent agreement with that derived  using the MAHs. This shows that
our  method  for  constructing  MAHs  is, within  the  EPS  framework,
self-consistent, and  does not  suffer from the  inconsistencies found
when using the LC93 `binary' method.

The right panel of Figure~\ref{fig:histogif} plots similar results but
now for the  6042 haloes in the mass range  $2.0 \times 10^{11} h^{-1}
\Msun  \leq M_{\rm  FOF} \leq  4.0 \times  10^{12} h^{-1}  \Msun$. The
agreement with the MAHs is somewhat poorer as in the case of the lower
mass  range, with  significantly  higher formation  redshifts for  the
haloes in the simulation.   This is, off course, another manifestation
of the inconsistencies  found between the MAHs in  the simulations and
those computed using the EPS formalism (cf., Figure~\ref{fig:gifmah}).

\subsection{Ellipsoidal collapse}
\label{sec:ellipsoid}

In  addition to  the inconsistencies  with the  MAHs  indicated above,
various  studies in  the  past  have pointed  out  that the  standard,
unconditional  PS  mass  function (equation~[\ref{PS}])  over  (under)
predicts  the  number of  low  (high)  mass  haloes when  compared  to
numerical simulations  (e.g., Jain \& Bertschinger  1994; Tormen 1998;
Gross \etal 1998; Governato \etal 1999). This is generally interpreted
as a consequence of the assumption of spherical collapse, and numerous
studies have shown that  considering ellipsoidal rather than spherical
collapse brings  the PS  mass function in  much better  agreement with
simulations (e.g., Monaco 1995; Bond \& Myers 1996; Audit, Teyssier \&
Alimi  1997; Lee  \& Shandarin  1998; Sheth  \& Tormen  1999; Lanzoni,
Mamon  \& Guiderdoni  2000; Sheth,  Mo \&  Tormen 2001;  Jenkins \etal
2001).

We can investigate a modification of  the spherical collapse model, by
multiplying the critical collapse  density  $\delta_c$ (as defined  in
Section~\ref{sec:background}) with a fudge factor $a$ and by examining
how  $a$ depends on halo  mass, if at all.   One of  the advantages of
using     the   scaled    variable    $\tilde{\omega}_f$     is   that
$P(\tilde{\omega}_f)$ is independent of   $a$, at least for   the MAHs
computed using  the EPS formalism.  In the  case of the haloes  in the
simulation one  has that  $\tilde{\omega}_f  \propto  a$, and one  can
therefore      immediately      re-scale   the    histograms        in
Figure~\ref{fig:histogif} to  determine  the  best-fit value  of  $a$.
Doing  so we find $a  \simeq 0.94$  and $a  \simeq  0.82$ for the mass
intervals  plotted in the panels  on the left ($M_0=8.4 \times 10^{11}
h^{-1} \Msun$) and    right ($M_0=1.1 \times  10^{13} h^{-1}  \Msun$),
respectively.   Similarly, we find  that for $a  \simeq  0.8$ the AMAH
plotted in Figure~\ref{fig:gifmah}  ($M_0=2.75 \times  10^{14}  h^{-1}
\Msun$) is  in    excellent agreement  with  the  MAHs   found in  the
simulation.    Thus it  seems  that  a  modification of  the spherical
collapse model whereby $\delta_c$  decreases with increasing halo mass
can bring the  EPS MAHs in  excellent agreement with  the simulations.
Interestingly, as shown by Sheth, Mo \& Tormen (2001), this is exactly
the kind of behavior one expects if one takes into account that haloes
are ellipsoidal rather than spherical. These authors obtain a modified
critical collapse overdensity given by
\begin{equation}
\label{ellips}
\delta_{ce}(M,z) = \delta_{c}(z) \left( 1 + 0.47 \left[{\sigma^2(M)
\over \delta^2_{c}(z)} \right]^{0.615}\right).
\end{equation}
Here $\delta_c(z)$  is the standard  value for the  spherical collapse
model. This  modification results in  halo mass functions that  are in
excellent agreement  with those found in simulations  (Sheth \& Tormen
1999; Jenkins \etal 2001).
\begin{figure*}
\centerline{\psfig{figure=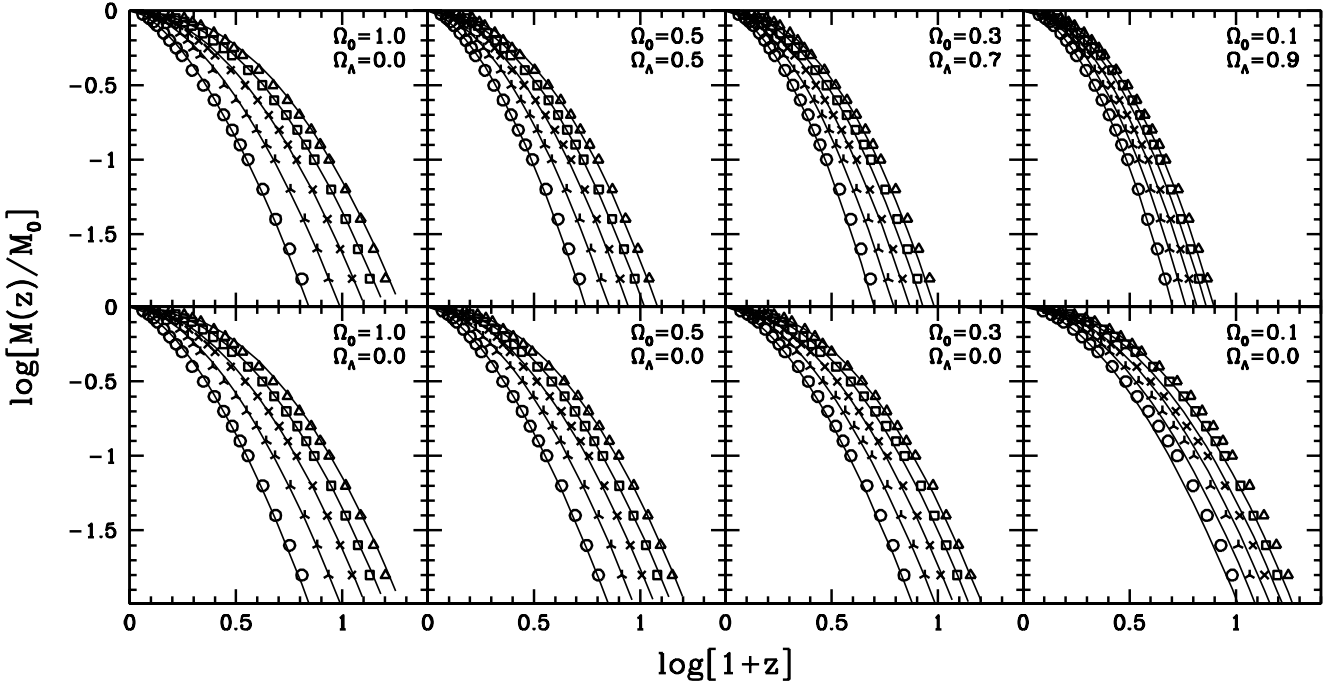,width=\hdsize}}
\caption{Average MAHs  for various  cosmologies (all with  $\sigma_8 =
1.0$ and  $h=0.65$). Results  are shown for  five masses:  $5.0 \times
10^9 h^{-1} \Msun$ (open triangles), $5.0 \times 10^{10} h^{-1} \Msun$
(open  squares), $5.0  \times  10^{11} h^{-1}  \Msun$ (crosses),  $5.0
\times  10^{12} h^{-1}  \Msun$ (tripods),  $5.0 \times  10^{13} h^{-1}
\Msun$ (open circles).  Averages are determined from $10^3$ random MAH
realizations.  Solid  lines correspond  to the universal  MAH computed
using the recipe given in  Appendix~A, and provide an excellent fit to
the average MAHs.}
\label{fig:allMAH}
\end{figure*}

Thus, two  separate statistics,  the unconditional halo  mass function
and the halo formation times (i.e., the MAHs), both suggest a critical
collapse density that increases  with decreasing halo mass.  Therefore
it  might  be  worthwhile  to  try and  incorporate  a  mass-dependent
critical collapse  density in the  EPS formalism. This  requires, what
pundits call, determining the  upcrossing statistics for a {\it moving
barrier}.   However,  as  discussed  by  Sheth  \&  Tormen  (2001)  no
analytical expression for  the conditional probability function (i.e.,
a moving barrier equivalent of equation~[\ref{probSS}]) is known for a
barrier of the form of  equation~(\ref{ellips}), and one has to resort
to  either  an  approximate  fitting  function,  or  one  has  to  use
time-consuming  Monte-Carlo simulations  to  determine the  upcrossing
statistics. Unfortunately,  as discussed in detail by  Sheth \& Tormen
(2001), neither of these  two methods are appropriate for constructing
merger trees.  Therefore, in what  follows we adhere to  the standard,
spherical collapse model,  but we caution the reader  that, taking the
numerical  simulations at face  value, the  MAHs thus  derived contain
slight, mass-dependent inaccuracies.

\section{A universal form for the mass accretion histories}
\label{sec:universal}

Figure~\ref{fig:allMAH} plots the average MAHs for various halo masses
(different symbols) and  cosmologies (different panels).  Upper panels
correspond    to   $\Lambda$CDM    cosmologies   with    $\Omega_0   +
\Omega_{\Lambda}  = 1$, while  lower panels  are for  OCDM cosmologies
without  cosmological constant (in  both cases  we adopt  $h=0.65$ and
$\sigma_8=1.0$).   These plots  clearly show  the  well-known behavior
that smaller mass haloes form earlier, which is a direct reflection of
the  fact  that  $\sigma(M)$  increases  with  decreasing  mass.   The
cosmology-dependence  of the MAHs  is easily  understood if  one takes
into account how  the mass variance $\sigma(M)$ and  the linear growth
factor $D(z)$ depend  on cosmology. On the mass  scales of interest, a
decrease  in  $\Omega_0$ causes  $\sigma(M)$  to  decrease (for  fixed
$\sigma_8$),  which implies  lower formation  redshifts  (i.e., slower
accretion rates). At the same time, a decrease in $\Omega_0$ causes an
increase in  $D(z)$ (at fixed redshift),  so that a  time step $\Delta
\omega$ implies a  larger $\Delta z$.  This drives  the MAHs to higher
formation  times.   The  net  result  of  a  decrease  in  $\Omega_0$,
therefore,  depends on which  of these  two effects  dominates.  Since
${\dd}\sigma/{\dd}M$ decreases  with $\Omega_0$ the $\sigma(M)$-effect
on $z_f$ is stronger for less massive systems.  Therefore, one expects
the $\sigma(M)$-effect to dominate  for small enough masses, resulting
in  a decrease  of $z_f$.   Furthermore, the  increase of  $D(z)$ with
decreasing    $\Omega_0$   is   stronger    for   a    Universe   with
$\Omega_{\Lambda}=0$  than  for  one  with  $\Omega_{\Lambda}  =  1  -
\Omega_0$, so  that the  mass scale below  which $z_f$  decreases with
decreasing $\Omega_0$  is higher  in an open  cosmology compared  to a
$\Lambda$-cosmology.  This  behavior is nicely reproduced  by the MAHs
plotted in  Figure~\ref{fig:allMAH}.  In the  upper panels, decreasing
$\Omega_0$ only  very mildly  affects the MAH  of a $5  \times 10^{13}
h^{-1} \Msun$  halo.  For  this mass the  two effects  mentioned above
largely   cancel   each   other.    For  less   massive   haloes   the
$\sigma(M)$-effect  dominates, causing  the  haloes to  form later  in
lower-$\Omega_0$ cosmologies.  In fact,  for $\Omega_0 = 0.1$ the mass
dependence of MAHs is much  reduced compared to the EdS cosmology.  In
OCDM cosmologies, the $D(z)$-effect  is relatively stronger and now it
are the low mass systems  that are hardly affected, while more massive
haloes increase their formation redshift with decreasing $\Omega_0$.

After experimenting with a variety  of fitting functions, we find that
the AMAHs are well fitted by the following simple form:
\begin{equation}
\label{unimah}
{\rm log} \langle \Psi(M_0,z)  \rangle = -0.301 \left[ {{\rm log}(1+z)
\over {\rm log}(1+z_f)} \right]^{\nu}
\end{equation}
where $z_f$ and $\nu$ are free fitting parameters. Note that with this
definition $z_f$  corresponds to the formation redshift  as defined in
Section~\ref{sec:simform} (i.e., $M(z_f) = M_0/2$). In what follows we
shall  refer  to   equation~(\ref{unimah})  as  the  `universal'  mass
accretion  history. Recently  Wechsler \etal  (2001)  investigated the
mass  accretion  histories  of  individual  dark matter  haloes  in  a
$\Lambda$CDM  simulation which  they fitted  with $\Psi(M_0,z)  = {\rm
e}^{-\alpha  z}$  (with  $\alpha$   a  free  fitting  parameter).   In
Appendix~B  we  compare this  one-parameter  fitting  function to  the
universal MAH of equation~(\ref{unimah}), and we show how $\alpha$ may
be estimated from the formation redshift $z_f$.
\begin{figure}
\centerline{\psfig{figure=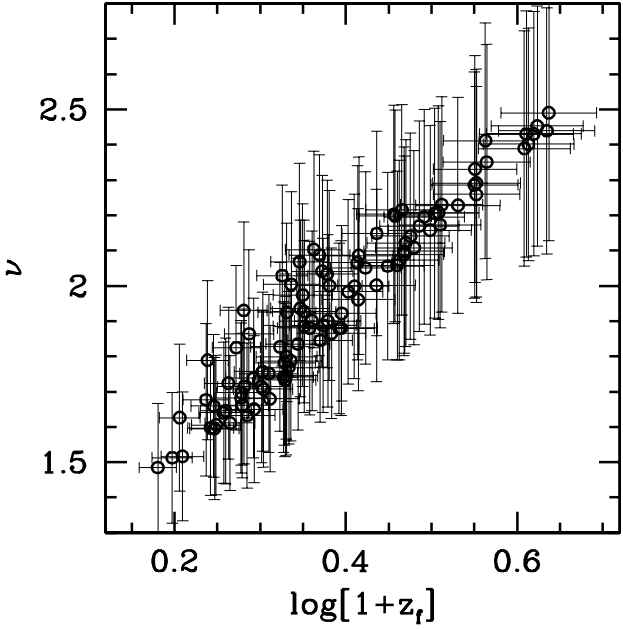,width=\hssize}}
\caption{The   correlation  between   the  best-fit   values   of  the
scale-parameters  $\nu$ and $z_f$  of the  universal MAH,  obtained by
fitting  equation~(\ref{unimah}) to  the AMAHs  of dark  matter haloes
with   a  large  range   of  masses   and  for   a  wide   variety  of
cosmologies. Errorbars correspond to the formal errors returned by the
fitting routine used.  Note  that $\nu$ and $\log(1+z_f)$ are strongly
correlated suggesting that a single parameter suffices to describe the
average   mass  accretion  history   of  a   dark  matter   halo  (see
Appendix~A).}
\label{fig:model}
\end{figure}

In  order to  investigate  how the  scale-parameters  $\nu$ and  $z_f$
depend on mass  and cosmology we proceed as  follows: we randomly draw
values for  $\Omega_0$, $M_0$, and $\sigma_8$ from  the intervals $0.1
\leq \Omega_0  \leq 1.0$, $10^{9}  h^{-1} \Msun \leq M_0  \leq 10^{14}
h^{-1} \Msun$  and $0.5  \leq \sigma_8 \leq  1.5$.  For each  of these
models we compute the AMAH from $10^3$ random realizations of the MAH,
and  we find  the best-fit  values of  $\nu$ and  $z_f$.  In  total we
construct 50  AMAHs with  $\Omega_{\Lambda} = 0$  and another  50 with
$\Omega_{\Lambda}    =     1    -    \Omega_0$.      As    shown    in
Figure~\ref{fig:model},  the best-fit  values of  $\nu$ and  $z_f$ are
strongly correlated,  suggesting that  a single parameter  suffices to
describe the  average MAHs.  In  Appendix~A we present  a step-by-step
recipe  for computing $\nu$  and $z_f$  as function  of halo  mass and
cosmology directly.

The solid lines in Figure~\ref{fig:allMAH} correspond to the universal
mass  accretion histories  with $\nu$  and $z_f$  computed  using this
recipe.    Only  in   the   extreme  case   with  $\Omega_0=0.1$   and
$\Omega_{\Lambda}=0.0$  (lower right  panel), does  the  universal MAH
fail  to accurately  fit the  average MAHs.  In all  other  cases, the
universal  MAH is in  excellent agreement  with average  MAHs obtained
using the method described in Section~\ref{sec:new}.

\section{Application: The Accretion Rate of Dark Matter Haloes}
\label{sec:appl}

Using the universal mass accretion history, the average rate at which 
haloes of mass $M_0$ accrete mass can be written as
\begin{equation}
\label{dmdt}
{{\dd}M \over {\dd}t}(z) = M_0 {{\dd}\Psi \over {\dd}t}(M_0,z)
\end{equation}
We  can use this  to compute the  total comoving dark matter accretion
rate of all haloes that at $z=0$  have masses between $M_1$ and $M_2$,
by weighting each   halo by the   present day, comoving  {\it  number}
density $n(M,z=0)$:
\begin{equation}
\label{mar}
{{\dd} \rho  \over  {\dd}t}(z;M_1,M_2)  =  \int_{M_1}^{M_2} {{\dd}\Psi
\over {\dd}t}(M,z)  \, { {\dd}n \over {\dd}  \, {\rm  ln} M}(M,z=0) \;
{\dd}M
\end{equation}
Here $\rho(z)$ is defined as the comoving mass density at redshift $z$
of all  main progenitor  haloes that by  $z=0$ have evolved  to become
haloes  with $M_1  \leq M_0  \leq M_2$.   Using the  universal  MAH of
equation~(\ref{unimah})    and    the     PS    mass    function    of
equation~({\ref{PS}), this integral is easily computed numerically.

If we make  the simplifying assumption that all  baryons inside haloes
with present day  masses in the interval $M_1 \leq  M_0 \leq M_2$ were
instantaneously  turned  into stars  the  moment  they were  accreted,
multiplying  equation~(\ref{mar}) with  the universal  baryon fraction
$f_{\rm    bar}$   gives    the   comoving    star    formation   rate
${\dd}\rho_*/{\dd}t$ as  function of  redshift. For $M_0  \gta 10^{13}
h^{-1} \Msun$  the cooling  time is longer  than the Hubble  time, and
such systems are therefore not expected to contribute significantly to
the cosmic star formation rate.   Haloes with $M_0 \lta 10^{10} h^{-1}
\Msun$  have virial  velocities  $V_{\rm  vir} \lta  30  \kms$, and  a
typical background UV radiation field can prevent the gas from cooling
(e.g., Babul \&  Rees 1992; Kepner, Babul \&  Spergel 1997). Therefore
it is  to be expected  that the majority  of star formation  occurs in
haloes  in this  mass  range.  We  therefore  set $M_1=10^{10}  h^{-1}
\Msun$   and    $M_2   =   10^{13}   h^{-1}    \Msun$,   and   compute
${\dd}\rho_*/{\dd}t$ using $f_{\rm  bar} = 0.019 \Omega_0^{-1} h^{-2}$
(Tytler \etal 1999).
\begin{figure*}
\centerline{\psfig{figure=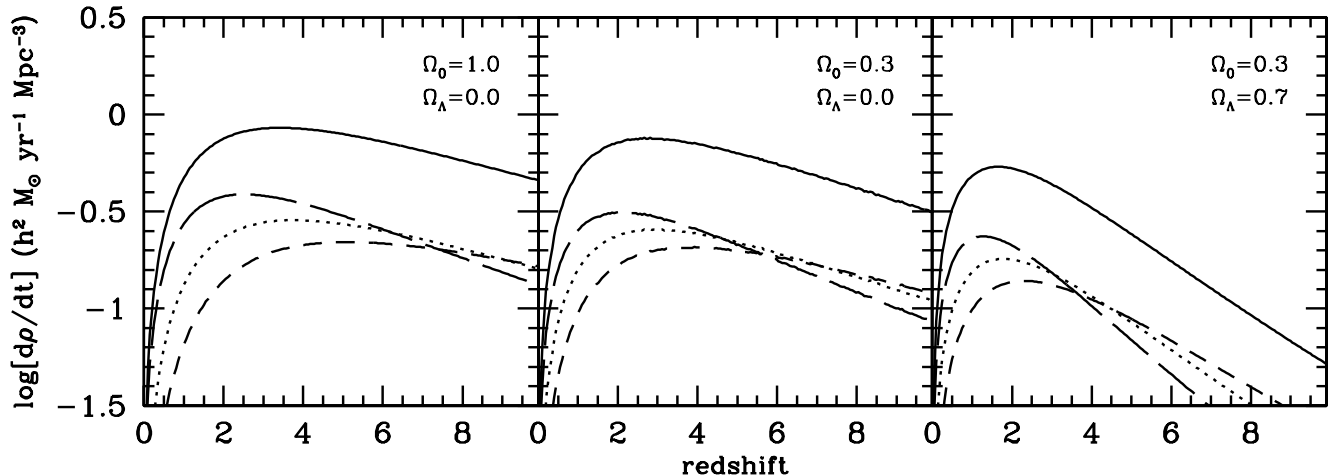,width=\hdsize}}
\caption{The    baryonic    mass    accretion   rates    defined    by
equation~(\ref{mar})  multiplied with  the  Universal baryon  fraction
$f_{\rm bar}$. Solid lines  correspond to the comoving accretion rates
integrated  over all haloes  with present  day masses  $10^{10} h^{-1}
\Msun \leq  M_0 \leq 10^{13} h^{-1}  \Msun$.  In addition  we plot the
contribution  to  these curves  from  three  separate mass  intervals:
$10^{10}   h^{-1}  \Msun   \leq   M_0  \leq   10^{11}  h^{-1}   \Msun$
(short-dashed  lines), $10^{11}  h^{-1}  \Msun \leq  M_0 \leq  10^{12}
h^{-1} \Msun$ (dotted lines), and  $10^{12} h^{-1} \Msun \leq M_0 \leq
10^{13} h^{-1} \Msun$ (long-dashed lines). Results are shown for three
different   cosmologies   as   indicated   (all  with   $h=0.65$   and
$\sigma_8=1.0$). The rapid increase at  low $z$ to a maximum accretion
rate at  $z \simeq 2$  is in good  agreement with observations  of the
cosmic star formation history (see discussion in text).}
\label{fig:mar}
\end{figure*}

The  results are  shown  in Figure~\ref{fig:mar}  for three  different
cosmologies, all with $h=0.65$  and $\sigma_8=1.0$ (solid lines).  The
short-dashed,  dotted,  and   long-dashed  curves  plot  the  separate
contributions from the  mass ranges $10 \leq {\rm  log}(M_0) \leq 11$,
$11 \leq  {\rm log}(M_0)  \leq 12$, and  $12 \leq {\rm  log}(M_0) \leq
13$,  respectively.   In  all three  cosmologies  ${\dd}\rho_*/{\dd}t$
increases rapidly at  low redshift, peaks in the  redshift interval $1
\lta z \lta  3$, and then declines (the steepness  of which depends on
cosmology).  This is in good agreement with observations of the cosmic
star formation  rate, which  increases by over  an order  of magnitude
from $z=0$  to $z=1$, and  which seems  to peak at  $1 \lta z  \lta 2$
(e.g., Lilly  \etal 1996;  Madau \etal 1996;  Steidel \etal  1996; and
references  therein).  Clearly  our assumption  of  instantaneous star
formation    with    $100$   percent    efficiency    is   a    severe
over-simplification.   In  reality,  there  will be  a  delay  between
accretion and  star formation  set by the  cooling and free  fall time
scales  of the halo.   In addition,  not all  baryons partake  in star
formation, as present  day gas mass fractions in  galaxies are clearly
not zero.   Furthermore, various  processes can temporarily  quench or
enhance  star formation  compared to  the cooling  rate,  and feedback
processes can  cause baryons to cycle through  multiple star formation
episodes. A  more elaborate comparison  with the observed  cosmic star
formation  rate will  have to  take  all these  effects into  account.
Nevertheless,  it is  reassuring that  our  oversimplified assumptions
already  yield   results  that  reproduce   the  main  characteristics
observed. This  suggests that indeed the baryonic  mass accretion rate
is the main  driving force for the cosmic  star formation history, and
that  the  universal MAH  derived  here may  proof  a  useful tool  in
modeling the history of star formation in the Universe.

\section{Conclusions}
\label{sec:concl}

We  have  presented  an  improved  method  for  determining  the  mass
accretion  histories  (MAHs)  of  dark  matter haloes,  based  on  the
$N$-branch merger-tree construction  algorithm of Somerville \& Kolatt
(1999). As we  have shown, this yields MAHs  with formation times that
are in excellent agreement with direct estimates based on the extended
PS  formalism. This  solves  an inconsistency  that hampered  previous
methods for  constructing MAHs,  which were based  on a  binary method
where  the assumption  was made  that the  most massive  progenitor is
always more massive  than half the mass of  the parent. This, however,
is a poor assumption and  results in halo formation redshifts that are
too high.

The MAHs and  halo formation times obtained using  our improved method
are   in   reasonable  agreement   with   high  resolution   numerical
simulations. The small discrepancies found  seem to be larger for more
massive haloes. Such mass dependence  is also found when comparing the
PS mass function with simulations. Various authors have suggested that
this  mass   function  discrepancy  can  be  solved   by  adopting  an
ellipsoidal collapse  model, rather than the  spherical collapse model
used in  the standard  PS formalism. Interestingly,  under ellipsoidal
collapse one predicts  the critical collapse density to  be higher for
less massive systems, which is  consistent with our mass dependence of
the halo formation time discrepancies found.

We  have shown  that the  average MAHs  of dark  matter haloes  have a
universal  functional form. The  dependence of  mass and  cosmology is
absorbed  by two  parameters  that can  be  computed accurately  using
simple fitting functions.  Together with the universal density profile
of  CDM haloes  (Navarro,  Frenk  \& White  1997),  and the  universal
angular momentum distribution of CDM haloes (Bullock \etal 2001), this
universal  mass accretion history  provides a  complete set  of simple
equations that can be used to model the structure and formation of the
population of dark matter haloes. The universal mass accretion history
is especially useful in modeling the formation of disk galaxies, where
the detailed effects of merging are not important (e.g., the models of
Firmani \& Avila-Reese  2000 and van den Bosch  2001a,b). In addition,
the  universal MAH allows  a straightforward  computation of  the mass
accretion rate of  dark matter haloes, which is  expected to drive the
cosmic  star   formation  rate,  and   allows  a  fast   but  accurate
investigation of  mass and/or cosmology dependencies  without the need
for constructing ensembles of actual mass accretion histories.


\section*{Acknowledgements}

I am  grateful to Anthony Brown,  Guinevere  Kauffmann, Houjun Mo, Adi
Nusser, Joel Primack, Rachel Somerville, Risa Wechsler and Simon White
for useful  discussions,   and to the  anonymous  referee  for helpful
comments.



\appendix

\section{A recipe for computing the average mass accretion history
of dark matter haloes}
\label{sec:AppA}

As shown in Section~\ref{sec:universal} the average MAH of dark matter
haloes is well fitted by the universal form:
\begin{equation}
\label{fitpsi}
{\rm log} \langle \Psi(M_0,z)  \rangle = -0.301 \left[ {{\rm log}(1+z)
\over {\rm log}(1+z_f)} \right]^{\nu}
\end{equation}
with $z_f$ and $\nu$ two scale-parameters that depend on halo mass and
cosmology,      and       which      are      strongly      correlated
(cf. Figure~\ref{fig:model}).
\begin{figure}
\centerline{\psfig{figure=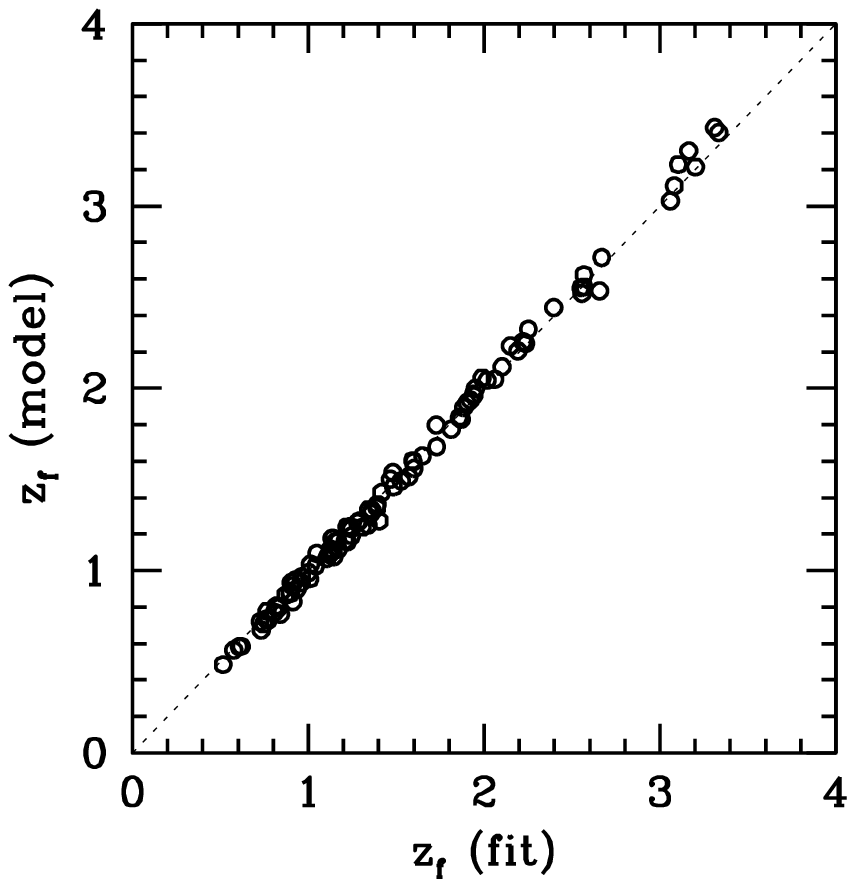,width=\hssize}}
\caption{The  values of  $z_f$ derived  from  equation~(\ref{zf}) with
$f=0.254$ versus the best-fit  values of $z_f$ determined from fitting
equation~(\ref{fitpsi}) to the average MAHs. Results are shown for 100
haloes    that   differ    in   both    mass   and    cosmology   (see
Section~\ref{sec:universal}).}
\label{fig:zzz}
\end{figure}

In order  to compute  $z_f$ directly from  EPS theory we  consider the
cumulative  probability   distribution  of  {\it   single  trajectory}
formation redshifts given by LC93:
\begin{equation}
\label{tformprob}
{\rm  P}(z >  z_f  \vert M_0)  =  {\rm erfc}  \left( {\delta_c(z_f)  -
\delta_c(0) \over \sqrt{2 [\sigma^2(f M_0) - \sigma^2(M_0)]}} \right)
\end{equation}
with  ${\rm erfc}(x)$  the complementary  error function  and $f=0.5$.
Here     $\delta_c(z)     =     \delta^0_{\rm
crit}[\Omega(z)]/D(z)$ with $D(z)$ the linear growth factor normalized
to unity at $z=0$ (see  Peebles 1980) and $\delta^0_{\rm crit}$ is the
critical threshold for spherical  collapse, which is well approximated
by
\begin{equation}
\label{delcrit}
\delta^0_{\rm crit}[\Omega(z)] = 0.15 (12 \pi)^{2/3} \Omega(z)^{p} 
\end{equation}
with
\begin{equation}
\label{p}
p = \left\{\begin{array}{ll}
0.0185 & \mbox{$\;\;\;$ if $\Omega_0<1$ and $\Omega_{\Lambda}=0$} \\
0.0055 & \mbox{$\;\;\;$ if $\Omega_0+\Omega_{\Lambda}=1$} \\
\end{array} 
\right.
\end{equation}
(e.g., Navarro, Frenk \& White 1997), and
\begin{equation}
\label{omegaz}
\Omega(z) = {\Omega_0 (1+z)^3 \over \Omega_{\Lambda} +
(1-\Omega_0-\Omega_{\Lambda}) (1+z)^2 + \Omega_0 (1+z)^3}
\end{equation}
The  median value  for $z_f$  of equation~(\ref{tformprob})  is easily
obtained by solving for the root of
\begin{equation}
\label{zf}
\delta_c(z_f) = \delta_c(0) + 0.477 \sqrt{2
[\sigma^2(f  M_0) - \sigma^2(M_0)]}
\end{equation}
As pointed out by LC93, this is  {\it not} the same as the median halo
formation  redshift,  as  it  does  not necessarily  follow  the  main
progenitor.  However, since  equation~(\ref{zf}) at least contains the
proper scaling with cosmological parameters, one might hope to be able
to use this  simple equation to model the best-fit  values of $z_f$ by
tuning  the  parameter $f$.   We  find  excellent  agreement with  the
best-fit     values     of     $z_f$    (determined     by     fitting
equation~(\ref{fitpsi})  to  the  average  MAHs)  for  $f=0.254$  (see
Figure~\ref{fig:zzz}).
\begin{figure*}
\centerline{\psfig{figure=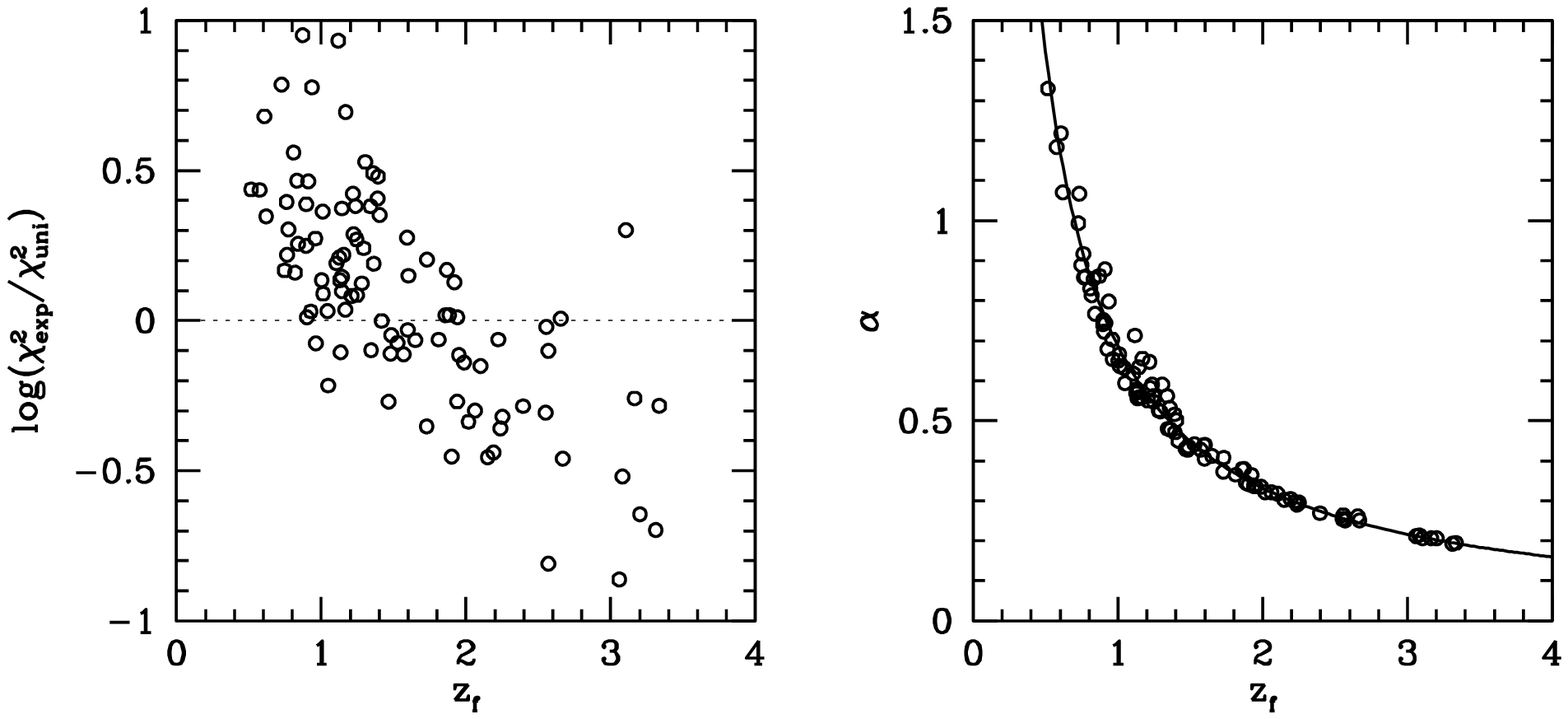,width=\hdsize}}
\caption{The  left panel  plots the  ratio of  $\chi^2_{\rm  exp}$ and
$\chi^2_{\rm  uni}$ as  function of  the best-fit  value of  $z_f$ for
$100$ haloes  with different masses and in  different cosmologies (see
Section~\ref{sec:universal}). As can  be seen, on average $\chi^2_{\rm
exp}  > \chi^2_{\rm  uni}$ for  $z_f <  1.5$, which  implies  that the
Universal  fitting  function  of  equation~(\ref{unimah})  provides  a
better  fit to  the  AMAH of  CDM  haloes that  form relatively  late,
whereas  the  exponential  MAH  of  equation~(\ref{expmah})  typically
provides a better fit  in cases with $z_f > 1.5$. In  the panel on the
right  we plot  the best-fit  values of  $\alpha$ versus  the best-fit
values  of  $z_f$, which  are  strongly  correlated.   The solid  line
corresponds to the  fitting function of equation~(\ref{relalp}), which
can be used  to convert $z_f$ computed using  the recipe in Appendix~A
to $\alpha$.}
\label{fig:chicomp}
\end{figure*}

Motivated  by the strong  correlation between  the best-fit  values of
$\nu$  and  $z_f$   we  find  that  the  power-index   $\nu$  is  well
approximated by
\begin{eqnarray}
\label{nuB}
\lefteqn{\nu = 1.211 + 1.858 \, {\rm log}[1+z_f] + 
0.308 \, \Omega_{\Lambda}^{2} - } \nonumber \\
& & 0.032 \, {\rm log}[M_0/(10^{11} h^{-1}  \Msun)]
\end{eqnarray}
(rms  error  of  $9.0  \times 10^{-4}$).   Equations~(\ref{zf})  (with
$f=0.254$) and~(\ref{nuB})  give the scale-parameters  $z_f$ and $\nu$
for any halo  mass and cosmology and therewith  completely specify the
universal mass accretion history~(\ref{fitpsi}).

Whenever we compute $\sigma(M)$ we use the following fitting function:
\begin{equation}
\label{sigmafit}
\sigma(M) = \sigma_8 \, {f(u) \over f(u_8)}.
\end{equation}
Here  $u_8  = 32  \Gamma$  (with  $\Gamma$  the power  spectrum  shape
parameter),
\begin{equation}
\label{ufit}
u   =   3.804   \times   10^{-4}   \,   \Gamma   \,   \left({M   \over
\Omega_0}\right)^{1/3},
\end{equation}
with $M$ in units of $h^{-1} \Msun$, and
\begin{eqnarray}
\label{fu}
\lefteqn{f(u) = 64.087 \left[ 1 + 1.074 \, u^{0.3} - 1.581 \, u^{0.4}
+ \right. } \nonumber \\
& & \left. 0.954 \, u^{0.5} - 0.185 \, u^{0.6} \right]^{-10}
\end{eqnarray}

This fitting function, which is  accurate to better than $0.5$ percent
over the  mass range $10^{6} h^{-1}  \Msun \leq M  \leq 10^{16} h^{-1}
\Msun$, is  only valid for  a spatial top  hat filter, and based  on a
power spectrum $P(k) = k \,  T^2(k)$ (i.e., we assume that the initial
power spectrum  has a Harrison-Zeldovich  form $P(k) \propto  k$) with
$T(k)$ the transfer function given by Bardeen \etal (1986):
\begin{eqnarray}
\label{transfer}
\lefteqn{T(k) = {{\rm ln}(1 + 2.34q) \over 2.34 q}
 \times} \nonumber \\
 & &  \left[ 1 + 3.89 q + (16.1 q)^2 + (5.46 q)^3 + (6.71 q)^4
 \right]^{-1/4} 
\end{eqnarray}
Here  $q = k/\Gamma$, with  $k$ in $h  {\rm Mpc}^{-1}$.

\section{An alternative form for the Universal mass accretion history}
\label{sec:AppB}

After this paper was  submitted Wechsler \etal (2001; hereafter WBPKD)
presented a similar investigation into the mass accretion histories of
CDM haloes.  They used the exponential form
\begin{equation}
\label{expmah}
\Psi(M_0,z) = {\rm e}^{-\alpha z}
\end{equation}
to  fit  the  mass  accretion  histories of  individual  haloes  in  a
$\Lambda$CDM  simulation.  There  are subtle  differences  between the
fitting function suggested here and  the one used by WBPKD, warranting
a closer  inspection which,  if any, provides  a better fit.   To that
extent  we have fitted  the AMAHs  of the  100 haloes  (with different
masses      and      different      cosmologies)     presented      in
Section~\ref{sec:universal}    with    the    exponential    MAH    of
equation~(\ref{expmah}).       In      the      left     panel      of
Figure~\ref{fig:chicomp}     we     plot    ${\rm     log}(\chi^2_{\rm
exp}/\chi^2_{\rm uni})$ as function of $z_f$.  Here $\chi^2_{\rm exp}$
and $\chi^2_{\rm  uni}$ correspond to the best-fit  values of $\chi^2$
for     the    fitting    functions     of    equations~(\ref{expmah})
and~(\ref{unimah}),   respectively.     As   can   be    seen,   ${\rm
log}(\chi^2_{\rm exp}/\chi^2_{\rm uni})$ and $z_f$ are fairly strongly
correlated to the  extent that haloes that form  relatively late ($z_f
\lta  1.5$)  are  on  average  better  fit by  the  Universal  MAH  of
equation~(\ref{unimah}), whereas the opposite  is true for haloes that
form early ($z_f \gta 1.5$). 

In the  right panel of  Figure~\ref{fig:chicomp} we plot  the best-fit
values of $\alpha$ as function of the best-fit values of $z_f$. As can
be seen  there is a  strong correlation between these  two parameters,
which is well fitted by
\begin{equation}
\label{relalp}
\alpha = \left( {z_f \over 1.43} \right)^{-1.05}
\end{equation}
(solid line). The fact that $\alpha$ and  $z_f$ are so well correlated
means  that   one can use   equation~(\ref{relalp})  and the recipe in
Appendix~A to estimate the value of $\alpha$ for  the average MAH of a
dark  matter halo of  arbitrary  mass and   cosmology.  This can,  for
example, be used to improve  accuracy by using equation~(\ref{expmah})
instead of~(\ref{unimah}) as an analytical description for the AMAH of
haloes with $z_f \lta 1.5$.

Another advantage of having a  direct way to compute $\alpha$ from the
recipe outlined in  Appendix~A is that WBPKD have  shown that $\alpha$
is strongly correlated with the  concentration of the final halo; more
concentrated haloes form on average  earlier.  WBPKD have shown that a
good fit  is obtained with  $c_{\rm vir} = 8.2/\alpha$,  where $c_{\rm
vir} =  r_{\rm vir}/r_s$ with $r_{\rm  vir}$ the virial  radius of the
halo, and $r_s$  the characteristic radius of the  NFW (Navarro, Frenk
\& White 1997)  halo density profile.  Using the  recipe in Appendix~A
and  equation~(\ref{relalp}) one  can thus  compute the  {\it average}
concentration of a dark matter halo of any mass and for any cosmology.
There are two small caveats  here.  First of all, the relation between
$c_{\rm  vir}$ and  $\alpha$ has  only been  tested for  one cosmology
(with    $\Omega_0=0.3$,     $\Omega_{\Lambda}=0.7$,    $h=0.7$    and
$\sigma_8=1.0$) and it remains to  be seen whether this also holds for
other  cosmologies.  However,  since  the results  of  WBPKD  strongly
suggest that the halo concentration  is set entirely by its MAH (i.e.,
the scatter in $c_{\rm vir}$ for  haloes of a fixed mass is consistent
with the scatter in MAHs), it seems likely that this will be the case.
Secondly,    the   relation    between   $z_f$    and    $\alpha$   of
equation~(\ref{relalp})  is based  on EPS  MAHs, whereas  the relation
between $c_{\rm vir}$ and $\alpha$ is based on MAHs fitted directly to
simulations.  Since the EPS  MAHs and those extracted from simulations
show some inconsistencies  (see Section~\ref{sec:nbody}) there will be
a systematic,  but small,  error in the  values of $c_{\rm  vir}$ thus
derived. Since the discrepancy  between EPS and simulations depends on
mass  and cosmology,  the same  applies  for the  amplitudes of  these
errors.   Nevertheless,  since  for  most  cases  the  error  will  be
relatively small compared to the scatter in $c_{\rm vir}$, this method
allows one to compute halo  concentrations for arbitrary halo mass and
cosmology to sufficient accuracy for most purposes.

\label{lastpage}

\end{document}